\documentclass[aps,reprint,apalike,assymb,twocolumn,10pt,superscriptaddress]{revtex4}
\usepackage{graphicx}       
\usepackage{amssymb}

\begin{document}

\title{Physics of RecA-mediated homologous recognition}

\author{Kevin Klapstein}
\affiliation{Dept. of Biomathematics, UCLA, Los Angeles, CA  90095-1766}

\author{Tom Chou}
\affiliation{Dept. of Biomathematics, UCLA, Los Angeles, CA  90095-1766}

\author{Robijn Bruinsma}
\affiliation{Dept. of Physics, UCLA, Los Angeles, CA  90095-1547}


Submitted to: Biophysical Journal, Jan. 05, 2004

\begin{abstract}
Most proteins involved in processing DNA accomplish their
activities as a monomer or as a component of a multimer containing a relatively
small number of other elements.  They generally act locally, binding to one or a few
small regions of the DNA substrate. Striking exceptions are the \textit{E. coli}
protein RecA and its homologues in other species, whose activities are associated
with homologous DNA recombination. The active form of RecA in DNA
recombination is a stiff nucleoprotein filament formed by RecA and DNA, within
which the DNA is extended by $50$\%. Invoking physical and geometrical ideas,
we show that the filamentary organization greatly enhances the rate of homologous
recognition while preventing the formation of topological traps originating from
multi-site recognition.

\end{abstract}

\maketitle


\vspace{2mm}

\noindent {\bf INTRODUCTION}
\vspace{1mm}

RecA is a $38$kD \textit{E. coli} protein which plays a key role both in DNA repair and in the
exchange of genetic material by promoting DNA strand exchange. RecA or a RecA homologue has
been found in every species in which it has been sought \cite{RoC97}. RecA mediated strand
exchange is important in maintenance of the genome and essential for sexual reproduction.  To
facilitate strand exchange, RecA polymerizes onto DNA to form a right-handed helical filament with
a diameter of $\sim 10$nm \cite{HeG89} and a repeat length of $6$ monomers per helical turn
\cite{YuE92}. Filaments form on both single- and double-stranded DNA (ssDNA and dsDNA)
molecules.  One form of the filament is the extended filament, formed with an ATP cofactor
\cite{HeG89,YuE92}.  Extended filaments are very stiff. The persistence length of the extended
filaments formed with ssDNA is $\xi_{\text{ssDNA}\cdot\text{RecA}}\simeq 860$nm \cite{HeS99},
about $16$ times that of dsDNA. The DNA within the extended filament is stretched by $50\%$
relative to B-form DNA.

\begin{figure}
\begin{center}
\includegraphics[height=4in]{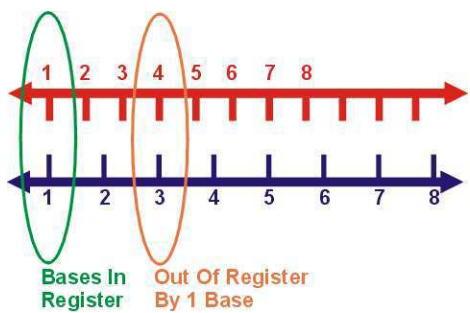}
\end{center}
\caption{Unequal spacing between base
pairs obstructs homologous alignment. Spacing between consecutive bases on
the bottom strand is 50\% larger than on the top strand, as for RecA coated
DNA. The DNA molecules are homologously aligned at the bases labeled $1$. By
base $3$ on the stretched molecule they are out of register by an entire
base.}
\label{Figure_01}
\end{figure}

Many different proteins with diverse functions act on DNA. Of these, RecA is rare
in that it forms stable protein {\it filaments} to accomplish its activity. Stretching the
DNA\ within the filament by $50\%$ is energetically expensive, requiring about
$\frac{2}{3}k_{B}T$ per base pair for dsDNA. This seems to present an
\emph{obstacle} to RecA activity by complicating the process of aligning regions
of homology between the two substrate DNA molecules.  Figure \ref{Figure_01}
shows how stretching one DNA molecule relative to the other makes it impossible
for the homologous regions of the two DNA substrates to stay ``in register'' with
each other. If the DNA molecules are homologously aligned at one base(pair), the
neighboring base(pair) is ``out of register'' by the difference in base(pair) spacing
between the stretched and unstretched DNA molecules. For RecA filaments, the
stretching is by a factor of $50$\%. Starting from a homologously aligned
base(pair), the next to neighboring base along the stretched molecule is already
out of register by an entire base.

Other DNA processing proteins, including those associated with DNA
recombination and repair, function without needing a filamentous structure,
yet RecA and its homologues have preserved the filament structure over 2.5
billion years of evolution in species as diverse as \textit{E. coli} and 
\textit{H. sapiens}. This strongly suggests that it is of value to the
system and necessary or at least advantageous in one or more of RecA's
activities. The aim of this paper is to use basic physical
considerations to understand the role of RecA filaments
in  RecA function.

\vspace{4mm}

\noindent {\bf FACILITATING INITIAL HOMOLOGOUS ALIGNMENT}

\vspace{2mm}

\noindent {\bf Incompatible Inter-Base Spacings}

\vspace{1mm}

\begin{figure}
\begin{center}
\includegraphics[height=3.0in]{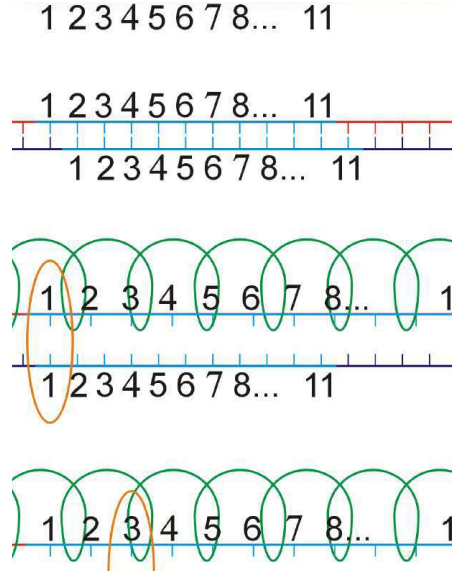}
\end{center}
\caption{Alignment between ssDNA
molecules. The molecules are homologous in the blue region. ($A$) Bases are a
distance $a_{0}$ apart in both molecules. All homologous bases are
aligned and in register. ($B$) The bottom strand has been moved to the right a
distance $a_{0}$. Now \emph{none} of the bases are aligned. ($C$) The
upper molecule is coated with RecA. Bases are spaced $3a_{0}/2$ apart.
Although the first base in the region of homology is aligned (orange circle), 
all others are out-of-register. ($D$) Moving the bottom strand to the
right a distance $a_{0}$ does not destroy the alignment, but merely
moves its location to the third base in the region of homology (orange
circle). All other bases remain out of register.}
\label{Figure_02}
\end{figure}

Homology recognition requires identification of many consecutive base pairs as
complementary and cannot be accomplished by identifying only a single pair of
complementary bases, yet the length of the region which can be simultaneously
compared is limited to a single base by the stretching of the DNA within the
filament. Surprisingly, we have found that stretching the DNA within the filament
does not impede but rather \emph{accelerates} the initial alignment of the
homologous regions of the DNA substrates. To see how, consider two B-form
DNA molecules having a region of homology. Denote the spacing between
consecutive base pairs of B-form DNA by $a_{0}$. If these molecules are parallel
then for some position of the one molecule relative to the other the homologous
regions will be aligned. As shown in figure \ref{Figure_02}$A$, every base pair
throughout the region of homology is then homologously aligned.  Figure
\ref{Figure_02}$B$ shows the effect of displacing one of the molecules a distance
$a_{0}$ relative to the other. Now, \emph{none} of the homologous base pairs are
aligned. When two DNA molecules have identical spacing between consecutive
bases, either \emph{all} bases in the region of homology are aligned or
\emph{none} of them are.

Now consider the case when one of the molecules is stretched. The spacing
between consecutive base pairs is $\eta a_{0}$ where the stretching factor $\eta
>1$, as shown with $\eta =\frac{3}{2}$ in figures \ref{Figure_02}$C$ and
\ref{Figure_02}$D$. This has two effects on the homologous alignment. First, a base
aligned with its homologous counterpart on the other molecule will form the
\emph{only} homologously aligned pair. The homologous alignment is lost almost
immediately as the two molecules get out of register. When one of the molecules is
shifted relative to the other by a distance $a_{0}$, however, a homologous
alignment is preserved. The base pair originally homologously aligned is now out
of alignment, but two nearby bases have moved into homologous alignment. The
behavior is analogous to the operation of a Vernier scale or a slide rule. Alignment
is now achieved at only one base pair at a time, but it is far more robust with
respect to the relative longitudinal position of the molecules than when $\eta $ is
the same for both molecules.

We define two substrates, $\mathcal{D}$ and $\mathcal{R}$. $\mathcal{D}$ is a very
long B-form dsDNA molecule containing a specific sequence $Aa$ of length $J$
base pairs. $\mathcal{R}$ is a RecA coated ssDNA molecule $J$ bases long and
consisting entirely of the sequence $A$. We first ask what steps must occur as
part of the process of identifying and aligning the $J$ base pair long region of
homology between these substrates.

\begin{figure}
\begin{center}
\includegraphics[height=4in]{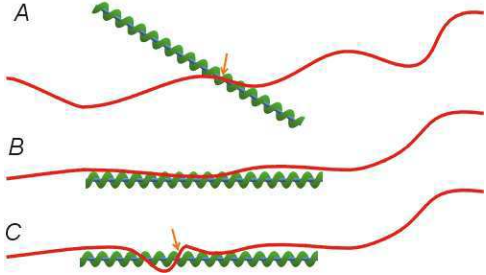}
\end{center}
\caption{Red curve, dsDNA molecule ($\mathcal{D}$); 
blue line, ssDNA, green helix, RecA filament ($\mathcal{R}$).
($A$) Contact at a point (orange arrow). An attractive interaction between
$\mathcal{D}$ and $\mathcal{R}$ exerts a torque around this point. ($B$) The
non-specific parallel orientation produced by the torque in A. ($C$) An
invading segment of $\mathcal{D}$ enters the RecA filament of $\mathcal{R}$
through the filament groove (orange arrow).}
\label{Figure_03}
\end{figure}

The first contact between the substrates will be as a result of diffusion,
and will occur at a point as shown in figure \ref{Figure_03}$A$. It has been
noted in the literature that there is a weak, non-specific (electrostatic)
attraction between the substrates \cite{KaB96}. This  exerts a torque on
the substrates, pulling them towards a loose, non-specific parallel
alignment as shown in figure \ref{Figure_03}$B$.

Recognizing a homologous alignment at any point now involves moving a short 
\emph{invading segment} of $\mathcal{D}$ into the \emph{interior position}
within the RecA filament of $\mathcal{R}$ through the groove in the RecA
helix, as shown in figure \ref{Figure_03}$C$. To keep the invading segment in
register with the DNA of $\mathcal{R}$ requires that the invading segment be
stretched by the stretching factor $\eta $ simultaneously with its entry into
the filament. Although this requires significant energy, the process as a
whole could remain energetically neutral if the binding energy of the
invading segment within the filament is of the same magnitude as the energy
required to stretch the invading segment. The movement of the invading
segment to the interior position would then be readily accessible to
thermal fluctuations and easily reversible. Correct homologous
alignment would stabilize the invading segment in the interior position by
the formation of hybrid base pairs, which would facilitate the movement of
more of $\mathcal{D}$ into the filament and further extension of the region of hybrid
DNA. In the absence of homologous alignment, this stabilization could not
occur and thermal fluctuations would remove the invading strand from the
interior position. In this way large sections of $\mathcal{D}$ and $\mathcal{%
R}$ can be quickly examined for homologous alignment by thermally driven
invasions of different regions of $\mathcal{D}$ at different points along $%
\mathcal{R}$, without the need for any relative diffusion of $\mathcal{D}$
and $\mathcal{R}$ in the longitudinal direction.

In a one-dimensional search, a simple model would be to wind a segment of $\mathcal{D}$
non-specifically into the filament and searches for homology by sliding
longitudinally within the filament. This is appealing because it avoids
repeatedly winding invading segments in and out of the filament, but it has
many problems. These include the fact that the experimental evidence
excludes significant  sliding during the search process \cite{Adz98}, consistent with the
observation that the longitudinal sliding of such large molecules would be
very slow and inefficient. This would also create a ``trap'' in which DNA
heterologous to the DNA in $\mathcal{R}$ acts as a suicide substrate by
winding into the filament, where it then blocks the entry of homologous DNA.
This poisoning of the reaction does not happen \emph{in vitro} and would be
lethal \emph{in vivo}. For these and other reasons, we reject this model.

\vspace{3mm}
\noindent {\bf Target Size Enhancement and Kinetic Perfection}

\vspace{2mm}
When $\mathcal{D}$ and $\mathcal{R}$ are parallel, movement of invading segments
between the interior and exterior positions can be very rapid, allowing
the entire parallel length of $\mathcal{D}$ and $\mathcal{R}$ to be efficiently
examined for homologous alignment. If there exists a point of alignment,
it will be found and the alignment extended throughout the entire  
homology. This reduces the problem to a question of how likely it is
that a random point of initial contact will result in a parallel orientation
which has a point of homologous alignment. Stretching of the DNA by RecA
plays a crucial role by enhancing this likelihood.

Define the \emph{target size} $\sigma$ for homologous alignment as the
range of longitudinal positions of one DNA molecule relative to another
which results in the homologous alignment of at least one base pair. For two
B-form DNA molecules, this requires great precision. The molecules must lie
within $\pm a_{0}/2$ of an exact alignment, so the target size is $\sigma
=a_{0}$. If they do lie in this range then {\it all} the
bases are aligned, but this is \emph{not} desirable. Later, we 
show that an extended region of homologous
alignment poses serious problems and is undesirable.

This contrasts with the situation between $\mathcal{D}$ and $\mathcal{R}$.
When $\mathcal{R}$ is stretched by a factor $\eta_{\mathcal{R}}$ the target
size for a region of homology $J$ base pairs long is found in Appendix A
to be

\begin{equation}
\sigma =\left[\left( J-1\right) \left(\eta_{\mathcal{R}}-1\right)
+1\right] a_{0}  
\label{Eqn_E_5_25}
\end{equation}

\noindent For two unstretched DNA molecules, $\eta _{\mathcal{R}}=1$ and equation 
\ref{Eqn_E_5_25} gives the expected result 

\begin{equation}
\sigma =a_{0}=0.34\text{nm}
\end{equation}

\noindent  but using the known RecA value of $\eta _{\mathcal{R}}=3/2$ gives 
$\sigma =(J+1)a_{0}/2$,
which scales linearly with the homology length. A modest $200$ base homology
gives 

\begin{equation}
\sigma \simeq 100a_{0}=34\text{nm}
\end{equation}

\noindent This huge target size, a $100$ fold increase relative to two unstretched DNA
molecules, offers an enormous advantage to the homology recognition process.
It is achieved only because of the stretching of the DNA within the RecA
filament.

The initial point of contact between $\mathcal{D}$ and $\mathcal{R}$
completely determines their relative position when parallel
(figures \ref{Figure_03}$A$ and $B$). Imagine holding this point fixed on 
$\mathcal{D}$ while changing it on the $\mathcal{R}$. Displacing the initial
point of contact on $\mathcal{D}$ by a distance $d$ will change the relative
position of $\mathcal{D}$ and $\mathcal{R}$ in the parallel arrangement by
an identical distance $d$. This means that to achieve a homologous alignment
the initial contact must occur within a region of length $\sigma $.

Knowing the target size allows us to estimate the reaction rate. In the
absence of sliding \cite{Adz98} the reaction rate cannot exceed the diffusion
limit, so the maximum ``on rate'' for the reaction is the Debye-Smoluchowski
rate. We approximate the cylindrical target of length $\sigma $ by a
spherical target of radius $\sigma /2$. We also estimate the diffusion
constant of $\mathcal{R}$, which is a cylinder of length $\ell$, by the
diffusion constant for a sphere with a radius of $\ell/2$. With these
approximations we find 

\begin{equation}
k_{\text{a}}\simeq \frac{2k_{B}T}{3\eta }\left( 1-\frac{1}{\eta _{\mathcal{R}}}
\right).
\label{KDIFF}
\end{equation}

This result is independent of the length of the homology, the number
of base pairs in the region of homology, and the spacing between consecutive
base(pair)s in unstretched B form DNA. It depends only on the temperature $T$, 
on $\eta$, the viscosity of the solution, and on the stretching factor $\eta_{\mathcal{R}}$. 
The stretching factor for DNA within the RecA filament is $\eta_{\mathcal{R}}=\frac{3}{2}$, 
and $\eta \simeq 10^{-3}$poiseuille at $20^{\circ }$C, so we have 
$k_{\text{a}}\simeq 5.3\times 10^{8}\text{ Mol}^{-1}\sec^{-1}$.

\vspace{5mm}

\noindent {\bf PREVENTING MULTIPLE HOMOLOGOUS ALIGNMENTS}

\vspace{2mm}

\noindent {\bf Topological Trapping}

\vspace{2mm}

\begin{figure}
\begin{center}
\includegraphics[height=4in]{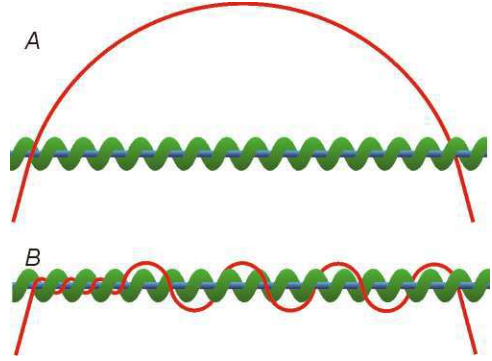}
\end{center}
\caption{Topological trapping resulting
from initiation of strand exchange at two separate points. Red curve = dsDNA
molecule ($\mathcal{D}$); blue line = ssDNA, green helix = RecA filament ($\mathcal{R}$). 
($A$) Homologous alignment at alignment regions $1$ (left) and $2$ (right). 
These are shown only at their initial point of closest approach
but may be of arbitrary length, {\it e.g.}, region $1$ may extend far to the left
of what is shown. ($B$) Extending alignment region $1$ to the right requires
that $\mathcal{D}$ be ``wound into'' the filament of $\mathcal{R}$. This
entails rotation of $\mathcal{D}$ and $\mathcal{R}$, forming ``counter
turns'' of $\mathcal{D}$ around $\mathcal{R}$. Because $\mathcal{D}$ is
fixed relative to $\mathcal{R}$ at region $2$, the counter turns are
``trapped'' between regions $1$ and $2$.}
\label{Figure_04}
\end{figure}

If homologous strand exchange between two substrates is initiated at two or
more separate points it will result in a problematic \emph{topological
trapping} of the reaction. To extend a region of hybrid DNA, at least one
strand of the external dsDNA must wind into the RecA filament \cite{HoR88}.
If exchange between the substrates is initiated at two separate points, this
motion produces compensating \emph{counter-turns} of the dsDNA around the
outside of the filament, as shown in figure \ref{Figure_04}. Extending the
hybrid DNA increases the number of counter turns while decreasing the length
of dsDNA which forms them, which rapidly decreases the radius of the counter-turns. 
This makes them very energetically expensive to produce, which
eventually stops extension of the hybrid DNA.

Topological trapping can occur only if homologous strand exchange is 
able to begin at two points. If initiation of homologous strand exchange
between the substrates at one point somehow prevents the initiation of
homologous strand exchange at any other point along the same two substrates,
topological trapping will never occur. To exploit this fact, the strand
exchange machinery must impart to the \emph{entire} length of the homology
the information that homologous strand exchange has been initiated between
them. The problem thus becomes finding a means of communicating over large
distances the fact that no further homologous strand exchange process should
be initiated.

We propose that a key function of the extended filament structure is to
prevent topological trapping. Both the stiffness of the filament and the
stretching of the DNA within it are essential to accomplishing this. Here,
we give a qualitative explanation of how this works. A more detailed
treatment appears in Appendix \ref{AppendixDblHitProb}.

Consider homologous exchange between $\mathcal{D}$ and $\mathcal{R}$ which
has extended to encompass a segment of the substrates which we call the 
\emph{first region}. Symmetry makes it sufficient to consider only the
sections of the substrates to one side of the first region. Number the
base(pair)s in ascending order to the right, beginning with the right most
base(pair) in the first region. If the second contact is at base $N$ on 
$\mathcal{D}$, it will be homologously aligned only if it at base pair $N$ on 
$\mathcal{R}$. Formation of a \emph{second region} produces a \emph{double
hit loop}, half of which is composed of ssDNA from $\mathcal{R}$ and the
remainder of dsDNA from $\mathcal{D}$. The second region is distinct if the
RecA filament passes through the double hit loop at least once, otherwise it
only extends the first region. We estimate the minimum work to form a double
hit loop for different $N$ and show that the resulting Boltzmann factor is
too small to permit the structure to form.

\begin{figure}
\begin{center}
\includegraphics[height=4.5in]{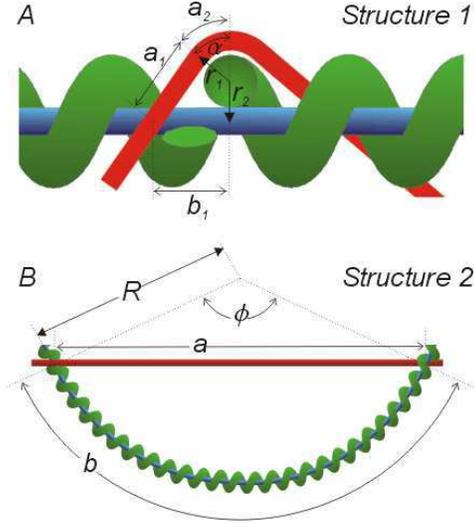}
\end{center}
\caption{Two homologous alignments
between the same substrates. Red curve = dsDNA molecule ($\mathcal{D}$); blue
line = ssDNA, green helix = RecA filament ($\mathcal{R}$). ($A$) Structure 
$1$ (small $N$). The helix is cut to allow clearer labeling of the figure.
The red and blue curves form the double hit loop. The RecA filament (green)
passes through this once. Because $\mathcal{R}$ is much stiffer than 
$\mathcal{D}$, we assume $\mathcal{R}$ does not bend significantly on this
length scale. ($B$) Structure $2$ (large $N$). Now the RecA filament
passes through the double hit loop many times. Because $\mathcal{R}$ is
longer here, we include bending with a radius of curvature $R$. We ignore
the bending of $\mathcal{D}$ necessary to enter the RecA filament.}
\label{Figure_05}
\end{figure}

When the RecA filament passes through the double hit loop $M=1$ times we
consider $N$ small. This forms \emph{structure }$1$, shown in figure 
\ref{Figure_05}$A$. As $\mathcal{R}$ is very stiff it will behave as a rigid rod
for small $N$, so we ignore any bending of $\mathcal{R}$. To align the 
$N^{\text{th}}$ base(pair)s of $\mathcal{D}$ and $\mathcal{R}$ then requires
stretching and bending $\mathcal{D}$. We ignore the work necessary to bend 
$\mathcal{D}$ and consider only the work needed to stretch it. This
underestimates the work, producing a lower bound.

\begin{figure}
\begin{center}
\includegraphics[width=3.6in]{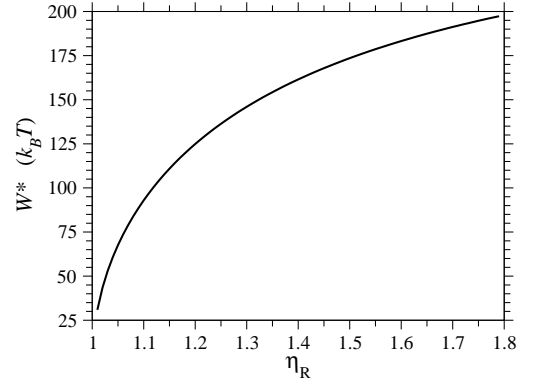}
\end{center}
\caption{The minimum work $W^*$ required to form structure $1$ as a
function of the stretching factor $\eta_{\mathcal{R}}$ of the DNA within substrate
$\mathcal{R}$ relative to B-form DNA.}
\label{Figure_06}
\end{figure}

The minimum work $W^{*}(\eta_{\mathcal{R}})$ required to form structure $1$ is calculated in
Appendix \ref{AppendixDblHitProb} and plotted as a function of the
stretching factor $\eta_{\mathcal{R}}$ in figure \ref{Figure_06}. $W^{*}$
increases with $\eta_{\mathcal{R}}$, making structure $1$ more difficult to
form. For RecA, $\eta_{\mathcal{R}}=1.5$ and the work is minimized by 
$\alpha^{*}=.84$ and $N^{*}\approx 25$. We call this the \emph{minimal} structure 
$1$. The energy of this is about $W^{*}\approx 170k_{B}T$. The
Boltzmann factor is $e^{-E_{i}/k_{B}T}$, so the probability of
forming structure $1$ is on the order of $e^{-170}$. Formation of a second
homologous alignment therefore does not occur for small $N$ as a result of
thermal fluctuations.

When the RecA filament passes through the double hit loop $M\geq 2$ times we
must include the bending of $\mathcal{R}$, but we ignore the bending of 
$\mathcal{D}$ as it enters or leaves the RecA filament. This shortens the
path for $\mathcal{D}$ and lengthens the radius of curvature for $\mathcal{R}$, 
both of which reduce the work, giving a lower limit on the work required
to achieve the second alignment. This is \emph{structure} $2$, shown in
figure \ref{Figure_05}$B$. The minimum work $W^{*}$ to form structure $2 $ is
calculated in Appendix \ref{AppendixDblHitProb}.

\begin{figure}
\begin{center}
\includegraphics[width=3.5in]{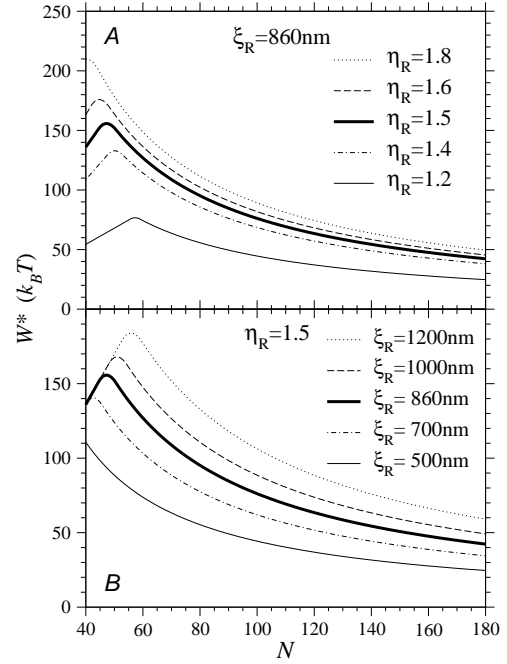}
\end{center}
\caption{$W^{*}$ as a function of $N$. 
($A$) For various stretching factors $\eta_{\mathcal{R}}$ 
with the physiological
persistence length of $\xi_{\mathcal{R}}=860$nm. The thick curve shows the
physiological case $\eta_{\mathcal{R}}=1.5$. ($B$) For various persistence lengths 
$\xi_{\mathcal{R}}$ 
with the physiological stretching factor of $\eta_{\mathcal{R}}=1.5$. The thick
curve shows the physiological case $\xi_{\mathcal{R}}=860$nm.}
\label{Figure_07}
\end{figure}

If $30\lesssim N\lesssim 100$, the expression for $W^{*}$ is complicated by the
division of the parameter space into different regions. 
Figures \ref{Figure_07}$A$ and $B$ are more informative. Figure \ref{Figure_07}$A$ shows 
$W^{*}(N)$ with the persistence length of $\mathcal{R}$ 
fixed at the physiological value of $\xi_{\mathcal{R}}=860$nm
for various values of $\eta_{\mathcal{R}}$. The curves all have a similar
shape, with the amplitudes increasing with increasing $\eta_{\mathcal{R}}$.
For small $N$, the minimal structure involves stretching $\mathcal{D}$ 
without significantly bending $\mathcal{R}$. In this case, 

\begin{equation}
W^{*} \approx F_{0}a_{0}\left( \eta _{\mathcal{D}}-1\right)N=F_{0}a_{0}\left(\eta _{\mathcal{R}}-1\right)N  
\label{Eqn_L_1_10}
\end{equation}

\noindent where $F_{0}$ is a known constant. Consequently, the left edges of these
curves in figure \ref{Figure_07}$A$ are spaced linearly in proportion to $\eta_{\mathcal{R}}$.
For the physiological value of $\eta_{\mathcal{R}}=1.5$, the minimal structure
at first involves only stretching $\mathcal{D}$ and $W^{*}$ forms a straight
line which increases with $N$ as in equation \ref{Eqn_L_1_10}. This
is because the larger $N$ is, the more base pairs in $\mathcal{D}$
must be stretched. As $N$ increases, the length of the segment of
$\mathcal{R}$ between regions $1$ and $2$ increases and the work required to
bend $\mathcal{R}$ decreases. At $N\simeq 48$, it becomes comparable
to the work required to stretch $\mathcal{D}$ and the minimal structure
becomes a combination of bending $\mathcal{R}$ and stretching $\mathcal{D}$.
As $N$ increases further, the work to bend $\mathcal{R}$ drops
further while the work to stretch $\mathcal{D}$ continues to increase so
bending $\mathcal{R}$ becomes a steadily larger part of the process. $W^{*}$
curves downward as this happens, and by $N\simeq 60$ the minimal
structure involves only bending $\mathcal{R}$ and no stretching $\mathcal{D}$. 
From here on, $W^{*}$ decreases when $N$ increases as $W^{*}\propto N^{-2}$.

The location of the transition from the stretching $\mathcal{D}$ regime with
$W^{*}\propto N$ to the bending $\mathcal{R}$ regime with $W^{*}\propto 
N^{-2}$ is influenced by $\eta_{\mathcal{R}}$. For larger $\eta _{\mathcal{R}}$, greater work is required to
sufficiently stretch a given number of base pairs, making $W^{*}$ larger for
larger values of $\eta _{\mathcal{R}}$. The work required to stretch
$\mathcal{D}$ thus becomes comparable to the work required to bend $\mathcal{R}$ at
smaller values of $N$, and the peak value of $W^{*}$ occurs at
smaller $N$ for larger values of $\eta _{\mathcal{R}}$.

Figure \ref{Figure_07}$B$ shows $W^{*}(N)$, with $\eta _{\mathcal{R}}$ fixed at the 
physiological value of $\eta _{\mathcal{R}}=1.5$, for various values of $\xi _{\mathcal{R}}$. The notable points here are
that increasing $\xi _{\mathcal{R}}$ makes it more difficult to bend $\mathcal{R}$, 
and therefore $W^{*}$ increases as $\xi _{\mathcal{R}}$ increases. This
also means that higher values of $\xi _{\mathcal{R}}$ push the transition from
the stretching $\mathcal{D}$ regime to the bending $\mathcal{R}$ regime to
higher values of $N$. For $\xi_{\mathcal{R}}=500$nm\ the transition
occurs at values of $N$ which are off the left side of figure \ref{Figure_07}$B$.

From these figures and from the calculation we see that for $N\lesssim 100$,
$W^{*}$ remains too high for structure $2$ to form as a result of thermal
fluctuations. This is only the case because of the large values of $\xi _{\mathcal{R}}$ 
and $\eta _{\mathcal{R}}$, since smaller values of either or both of
these decrease $W^{*}$ and make the structure more accessible to random
thermal processes. Since our calculation has produced only a lower limit on $W^{*}$ 
we can be confident that this conclusion is valid for values of $N$
which are at least this large.

For $N\gtrsim 60$, the minimal structure is dominated by
bending $\mathcal{R}$, and we can ignore stretching of $\mathcal{D}$. By
contrast, we can no longer ignore the work required to separate $\mathcal{D}$
and $\mathcal{R}$ against the non-specific attractive force which initially
brought them into alignment. This work will be proportional to the length of
the substrates between regions one and two. We use $\varepsilon $ for the
constant of proportionality. The work to form structure $2$ in this case is
calculated in appendix \ref{AppendixDblHitProb}. The function is found to
have a minimum with respect to $N$ at 

\begin{equation}
N^{*}=\frac{2}{a_{0}\eta _{\mathcal{R}}}\sqrt{\left( \frac{\xi _{\mathcal{R}}}{\varepsilon }\right) \left( 5-\sqrt{\frac{30}{\eta _{\mathcal{R}}}-5}\right) } +1   
\label{Eqn_XN_10}
\end{equation}

\begin{figure}
\begin{center}
\includegraphics[width=3.5in]{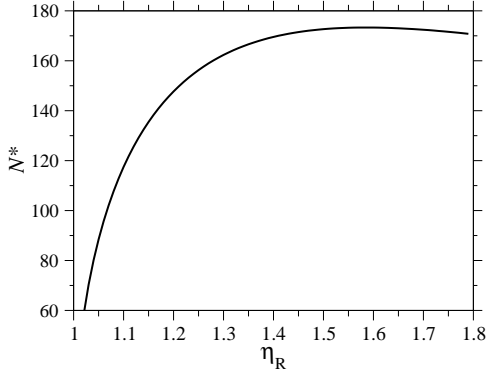}
\end{center}
\caption{$N^{*}$, the value of $N$ at which the energy
required to form structure $2$ is minimized, as a function of the stretching
factor $\eta_{\mathcal{R}}$. Note that $N^{*}$ has a maximum with
respect to $\eta_{\mathcal{R}}$ at $\eta_{\mathcal{R}}\simeq 1.58$, very near
the physiological value of $\eta_{\mathcal{R}}\simeq 1.5$.}
\label{Figure_08}
\end{figure}

\noindent This value for $N$ produces the minimal structure $2$ for large $N$.  The behavior of
$N^{*}$ in equation \ref{Eqn_XN_10} with respect to $\eta _{\mathcal{R}}$, as shown in figure
\ref{Figure_08}.  $N^{*}$ has a maximum with respect to $\eta _{\mathcal{R}}$. This occurs at
some value $\eta _{\mathcal{R}}^{*}$ which is found in the appendix to be
$\eta_{\mathcal{R}}^{*}\simeq 1.58$, strikingly close to the physiological value of
$\eta_{\mathcal{R}}=1.5$. This value maximizes the distance between the first and second regions
for which the work required to form structure $2$ is minimized.

\begin{figure}
\begin{center}
\includegraphics[width=3.5in]{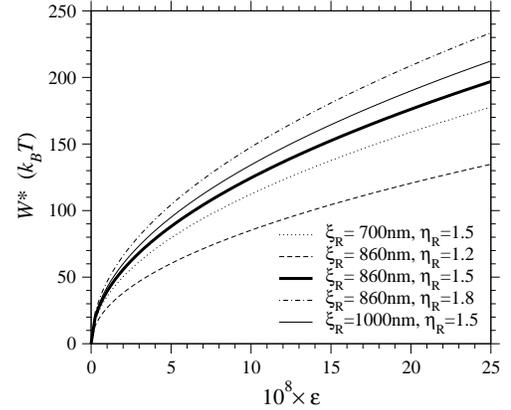}
\end{center}
\caption{The minimum work $W^{*}$ to form
structure $2$ for $N>60$ as a function of the constant of proportionality $\varepsilon$ 
for various stretching factors $\eta_{\mathcal{R}}$ and  persistence lengths 
$\xi_{\mathcal{R}}$. The thick curve shows the physiological 
case $\eta_{\mathcal{R}}= 1.5$ and $\xi_{\mathcal{R}}=860$nm.}
\label{Figure_09}
\end{figure}

\noindent Upon using $N^{*}$ in the expression for work gives the minimum value 

\begin{equation}
W^{*}=4k_{B}T\sqrt{\left(\xi_{\mathcal{R}}\varepsilon \right) \left( 5-\sqrt{\frac{30}
{\eta _{\mathcal{R}}}-5}\right) },
\label{Eqn_XN_12}
\end{equation}

\noindent shown in figure \ref{Figure_09}. The work $W^{*}$ increases as the
square root of $\varepsilon $. For the physiological values of $\eta _{\mathcal{R}}$ 
and $\xi _{\mathcal{R}}$, $W^{*}$ already reaches $\sim 50k_{\text{B}}T$
by $\varepsilon \simeq 0.2$nm$^{-1}$, so structure $2$ will not form as a
result of thermal motions for the physiological values of $\eta _{\mathcal{R}}$
and $\xi_{\mathcal{R}}$ when $\varepsilon \gtrsim 0.2$nm$^{-1}$. $W^{*}$
also increases as the square root of $\xi_{\mathcal{R}}$, and increases with
$\eta_{\mathcal{R}}$ in a more complicated fashion. Sufficiently small values
of $\eta_{\mathcal{R}}$ or $\xi _{\mathcal{R}}$ would produce values of $W^{*}$
which would be more accessible to thermal energies.

The probability that a second, local region of homology  aligns
$N$ base pairs away is $\propto N^{-6}$.
This small probability forms an  ``entropic'' 
barrier to formation of a second region of homology.
However, since $\eta _{\mathcal{R}}\simeq \eta _{\mathcal{R}}^{*}$, the value of $N$ 
for which the energetic obstacle to alignment is smallest, $N^{*}$, 
is made as large as possible, maximizing the entropic obstacle to
alignment. The two processes are ``tuned'' to work in a complementary
fashion, providing a further form of selection pressure for the maintenance
of stretching of the DNA by the RecA filament and, specifically, to a value
close to the physiological value of $\eta _{\mathcal{R}}=1.5$.

\vspace{6mm}

\noindent {\bf SUMMARY AND CONCLUSION}

\vspace{2mm}

RecA facilitated homologous recombination derives two advantages from the
stiff extended filament and the stretching of the DNA within the filament.

The first is a great increase in the efficiency of the homology search and
recognition process. This is a consequence of the increase in $\sigma$, the
target size for homologous alignment between the substrates. For a region of
homology $J$ bases in length stretched by a factor $\eta_{\mathcal{R}}$ we
have 

\begin{equation}
\sigma =\left[\left( J-1\right) \left( \eta _{\mathcal{R}}-1\right)
+1\right] a_{0}
\end{equation}

\noindent where $a_{0}$ is the spacing of B form DNA. For $\eta_{\mathcal{R}}\simeq 3/2$
this is $\sigma =(J+1)a_{0}/2$. This huge $\sigma $ allows large segments of the substrates
to be checked for homology without the need for sliding.

The second advantage is the prevention of topological trapping. Molecules
undergoing homologous strand exchange are kept in close proximity by the
region of hybrid DNA being formed. This greatly enhances the probability
that they will contact each other at additional points. Without the
stretching of the DNA within the extended filament, these secondary contacts
would often be in homologous alignment and capable of initiating a second
homologous strand exchange reaction. This would lead to a trapped state in
which a region of counter-wound DNA is trapped between two regions of hybrid
DNA, preventing completion of the exchange reaction.

The extended filament prevents homologous alignment at secondary contacts. 
Homologously aligned secondary contacts can only form through some
combination of stretching the DNA external to the filament and bending the
filament itself. For moderate distances from the point at which homologous
exchange is occurring, thermal fluctuations are incapable of sufficiently bending or
stretching the filament for a second hit to occur.  For larger distances, thermal
fluctuations are also unlikely to separate the locally aligned strands held together
by nonspecific electrostatic forces.  The interplay between  stiffness of the
filament and the stretching of the DNA within it ensures that homologous strand
exchange between two substrate molecules is initiated at only one point.

Both the stiffness of the filament and the extension of the DNA within it are necessary
features of the recombination apparatus. Without them, locating and aligning regions of
homology between two DNA molecules would be a slow and inefficient process, and the
exchange reaction would be prone to topological traps which would prevent completion
of the reaction and resolution of the products. These effects provide a selection pressure
to preserve the extended filament as a feature of homologous DNA recombination
facilitated by RecA and its homologues.

KK and TC acknowledge support from the National Science Foundation through
grant DMS-0206733. KK and RB acknowledge support from the UCLA Dean's
Funds.

\begin{appendix}

\section{Kinetic Perfection in the Homology Search}\label{AppendixKinPerf}

Consider $\mathcal{D}$ and $\mathcal{R}$ as defined in the text. The center
to center base(pair) spacing is $a_{0}$ in $\mathcal{D}$ and
$\eta_{\mathcal{R}}a_{0}$ in $\mathcal{R}$, so an $N$ base pair segment has length

\begin{eqnarray}
L_{\mathcal{D}} &=&\left( N-1\right) a_{0} \\
L_{\mathcal{R}} &=&\left( N-1\right) \eta _{\mathcal{R}}a_{0}
\end{eqnarray}

\noindent If $\mathcal{D}$ and $\mathcal{R}$ intersect at a point and rotate around
this point until parallel, at most one base within $\mathcal{R}$ and a base
pair in $\mathcal{D}$ will be homologously aligned. We wish to determine if
this parallel orientation produces an alignment.

To define ``aligned'' we consider only with the longitudinal positions of 
$\mathcal{D}$ and $\mathcal{R}$. A base on $\mathcal{R}$ is aligned with
whichever base pair on $\mathcal{D}$ is closest to it in the longitudinal
direction. When the aligned base and base pair are homologous they are
\emph{homologously aligned}.

To quantify this, first note that the center to center distance between
consecutive base pairs of $\mathcal{D}$ is $a_{0}$. Let the longitudinal
distance between the centers of the $k^{\text{th}}$ base pair on $\mathcal{D}$
 and the $l^{\text{th}}$ base on $\mathcal{R}$ be $\delta _{k,l}$. These
are aligned if 

\begin{equation}
\left| \delta_{k,l}\right| < {a_{0}\over 2}
\end{equation}

Denote the positions of initial contact by $x_{\mathcal{D}}$ along $\mathcal{D}$
and by $x_{\mathcal{R}}$ along $\mathcal{R}$. The parallel orientation is
achieved by rotating around $x_{\mathcal{D}}$ and $x_{\mathcal{R}}$, so these
completely determine the relative positions of $\mathcal{D}$ and $\mathcal{R}
$ once they are parallel. It is sufficient to fix $x_{\mathcal{R}}$ and ask
what values of $x_{\mathcal{D}}$ produce a homologous alignment in the parallel
orientation. If the position $x_{\mathcal{D}}$ is moved a distance $d$ along 
$\mathcal{D}$, $\mathcal{D}$ will be displaced relative to $\mathcal{R}$ by
this same distance $d$ in the parallel orientation. The range of $x_{\mathcal{D}}$ 
which produces a homologous alignment is therefore the same as the range
of longitudinal positions of $\mathcal{D}$ relative to $\mathcal{R}$ which
will produce such an alignment.

\begin{figure}
\begin{center}
\includegraphics[width=3.5in]{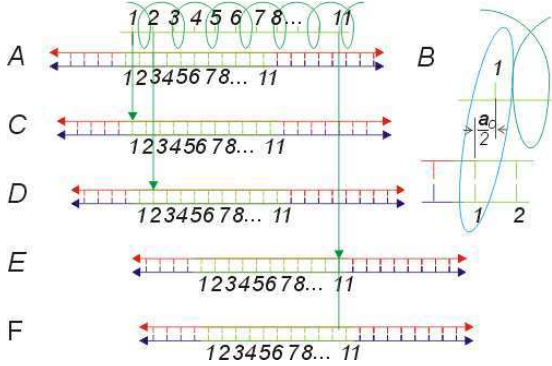}
\end{center}
\caption{Dark green helix,
RecA filament, light green line inside helix, ssDNA ($\mathcal{R}$); red and
blue lines with light green segments, dsDNA($\mathcal{D}$). $\mathcal{R}$ is
arbitrarily chosen to be $11$ bases in length. The light green region of 
$\mathcal{D}$ is homologous to $\mathcal{R}$. ($A$) $\mathcal{D}$ is as far to
the left as possible while maintaining a homologous alignment between 
$\mathcal{D}$ and $\mathcal{R}$ (between base and base pair $1$). ($B$) Close up
of the homologously aligned base and base pair in A. ($C$) Effect of shifting 
$\mathcal{D}$ to the right a distance $a_{0}/2$. ($D$) Effect of shifting
$\mathcal{D}$ further to the right, this time by a distance $a_{0}$.
($E$) Effect of $9$ more consecutive shifts of $\mathcal{D}$ to the right by a
distance $a_{0}$. ($F$) Effect of a final shift of $\mathcal{D}$ to the
right by a distance $a_{0}/2$.}
\label{Figure_10}
\end{figure}

Consider $\mathcal{D}$ and $\mathcal{R}$ as shown in figure \ref{Figure_10}$A$. 
Number the base pairs beginning with $1$ at the left most base pair in
the region of homology. Here, $\mathcal{D}$ is as far to the left as
possible while maintaining a homologous alignment between $\mathcal{D}$ and 
$\mathcal{R}$. The alignment is between base and base pair $1$, and the
center of base pair $1$ is $a_{0}/2$ to the left of
the center of base $1$. Figure \ref{Figure_10}$B$ shows an enlarged view of
this.

Moving $\mathcal{D}$ to the right by $a_{0}/2$ gives figure \ref{Figure_10}$C$, where the
centers of base pair $1$ and base $1$ are exactly aligned. Moving $\mathcal{D}$ to the
right by $\left( \eta _{\mathcal{R}}-1\right) a_{0}$ then produces figure \ref{Figure_10}$D$, in
which the centers of base pair $2$ and base $2$ are exactly aligned. Each displacement of
$\mathcal{D}$ to the right by $\left( \eta _{\mathcal{R}}-1\right) a_{0}$ now increments by
$1$ the base and base pair whose centers are exactly aligned. Starting with figure
\ref{Figure_10}$C$ and repeating this motion $(J-1)$ times produces figure \ref{Figure_10}$E$,
in which the centers of base pair $J$ and base $J$ are exactly aligned. A final movement
of $\mathcal{D}$ to the right by $a_{0}/2$ produces figure
\ref{Figure_10}$F$, where $\mathcal{D}$ is as far to the right relative to $\mathcal{R}$ as is
possible while still maintaining a homologous alignment between them.

The \emph{target size} $\sigma $ is range of longitudinal positions of $\mathcal{D}$ relative
to $\mathcal{R}$ which produces a homologous alignment between them. This is the
change in the position of $\mathcal{D}$ in going from figure \ref{Figure_10}$A$ to figure
\ref{Figure_10}$F$, which is given in eq. 1. If $\eta_{\mathcal{R}}$ has the known RecA value
of $\eta_{\mathcal{R}}=3/2$ we get $\sigma = (J+1)a_{0}/2$, which scales as the length of
the region of homology.

From $\sigma $ we can estimate the reaction rate. With no sliding\cite{Adz98}, diffusion
limits the maximum ``on rate'' $k_{\mbox{a}}$ for the reaction to the Debye-Smoluchowski
rate. Our target is cylindrical, but the magnitude should be reasonably approximated if we
substitute our target size for the diameter of a spherical target, $r\rightarrow \sigma /2$,
giving $k_{\text{a}}\simeq 2\pi D_{3}\sigma$, where $D_{3}$ is the three dimensional
diffusion constant. The length of $\mathcal{R}$ is 

\begin{equation}
\ell=\left(J-1\right) \eta _{\mathcal{R}}a_{0}\simeq J\eta _{\mathcal{R}}a_{0}
\end{equation}

\noindent Upon substituting this for the diameter of a spherical molecule, $r\rightarrow
\ell/2$, the three dimensional diffusion constant in a solvent with viscosity $\eta$, 
becomes $D_{3}=k_{B}T/(3\pi \eta \ell)$.
Assuming  $J$ is reasonably large, $\sigma \simeq
J\left(\eta _{\mathcal{R}}-1\right) a_{0}$, and 
$k_{a}$ is given by eq. \ref{KDIFF}.


\section{Double Hit Probability in the RecA Recombination System}
\label{AppendixDblHitProb}

If $N$ is small we consider structure $1$ as shown in Figure \ref{Figure_05}$A$. 
Define $\eta _{\mathcal{D}}$ and $\eta _{\mathcal{R}}$ as the
extension of $\mathcal{D}$ and $\mathcal{R}$ relative to the length of 
B-form DNA. Let $a_{0}$ be the spacing of base pairs in B form DNA and denote
the persistence lengths of $\mathcal{D}$ and $\mathcal{R}$ by $\xi _{\mathcal{D}}$ 
and $\xi _{\mathcal{R}}$ respectively. These have the numerical values 
$a_{0}=0.34$nm, $\xi _{\mathcal{D}}\simeq 53$nm, and $\xi _{\mathcal{R}}\simeq 860$nm. 
We model the RecA filament as a cylinder of radius $r_{\text{RecA}}$
whose axis follows a helical path of radius $r_{2}\simeq 3$nm. Treating $\mathcal{D}$ 
as a cylinder of radius $r_{\mathcal{D}}$ , the closest
approach of the center of $\mathcal{D}$ to the center of $\mathcal{R}$ will
be the sum of their radii, which we denote by $r_{1}=r_{\text{RecA}}
+r_{\mathcal{D}}\simeq 2.5$nm.

We wish to determine the minimum work required to form structure $1$. We
ignore bending of $\mathcal{R}$ on this length scale and also ignore the
work required to bend $\mathcal{D}$. We calculate the work solely from the
stretching of $\mathcal{D}$, making our calculation of the work a lower
bound.

The angle $\alpha $ can be varied to find the minimal-energy form of
structure $1$, the form produced with the minimum possible work, subject to
two constrains. Steric hinderance between the RecA filament and $\mathcal{D}$
at the points where it enters the RecA filament requires $\alpha \geq \pi/5$, while 
other physical considerations show that the minimum energy can occur only for 
$\alpha \leq \pi/2$.

Both regions can simultaneously be in homologous alignment only if $N$
satisfies 

\begin{equation}
\begin{array}{l}
2\left(a_{1}+a_{2}\right) =a_{0}\eta _{\mathcal{D}}\left( N-1\right) \\[13pt]
2b_{1}=a_{0}\eta _{\mathcal{R}}\left( N-1\right). 
\end{array}
\label{Eqn_A_0_11}
\end{equation}

\noindent Since $a_{2}=r_{1}\alpha $, trigonometry demands

\begin{equation}
a_{1}=\frac{r_{1}\cos \alpha +r_{2}}{\sin \alpha } \quad \mbox{and}\quad 
b_{1}=\frac{r_{2}}{\tan \alpha }+\frac{r_{1}}{\sin
\alpha }  \label{Eqn_A_0_23}
\end{equation}

\noindent Using  equations \ref{Eqn_A_0_11} and \ref{Eqn_A_0_23} we find

\begin{equation}
\begin{array}{l}
\displaystyle \left(N-1\right) =\frac{2}{a_{0}\eta _{\mathcal{R}}}
\left( \frac{r_{1}+r_{2}\cos \alpha  }{\sin \alpha }\right) \\[13pt]
\displaystyle \eta _{\mathcal{D}}=\frac{\eta _{\mathcal{R}}\left( r_{1}\alpha \sin \left( \alpha
\right) +r_{1}\cos \left( \alpha \right) +r_{2}\right)}{r_{1}+r_{2}\cos\alpha}  
\end{array}
\label{Eqn_A_0_39}
\end{equation}

\noindent We make the simplification of assuming that the force to stretch a dsDNA is 

\begin{equation}
F_{\mathcal{D}}=\left\{ 
\begin{array}{lll}
0 & \text{for} & \eta_{\mathcal{D}}<1 \\ 
F_{0} & \text{for} & \eta _{\mathcal{D}}\geq 1
\end{array}
\right.
\label{Eqn_A_0_40}
\end{equation}

\noindent Here, $F_{0}\approx 20k_{B}T/$nm, but we leave this parameter free for
the present.

Above $\eta_{\mathcal{D}}\approx 1.8$ equation \ref{Eqn_A_0_40} in not valid.
Here, the dsDNA melts while the force required to stretch it rises rapidly.
Further stretching breaks the sugar phosphate backbones of the DNA strands.
While our calculation may produce values of $\eta _{\mathcal{D}}>1.8$, we are
not concerned. We only wish to show that the minimal form of structure $1$
does not form as a result of random thermal fluctuations, and our
calculation will still accomplish this.

Using equation \ref{Eqn_A_0_40} for $F_{\mathcal{D}}$, the work required to
stretch the dsDNA to a final extension $\eta _{\mathcal{D}}$ (in units of 
$k_{B}T$) is 
\[
W_{\mathcal{D}}=\left\{ 
\begin{array}{lll}
0 & \text{for} & \eta _{\mathcal{D}}<1 \\ 
F_{0}\left(\eta_{\mathcal{D}}-1\right) a_{0}\left(
N-1\right) & \text{for} & \eta _{\mathcal{D}}\geq 1
\end{array}
\right. 
\]

\noindent Under our approximations, this is the only contribution to the total work.
Using equations \ref{Eqn_A_0_39} $W$ becomes a
function of the single variable $\alpha$

\begin{widetext}
\begin{equation}
W\left( \alpha \right) =\left\{ 
\begin{array}{lll}
0 & \displaystyle \text{for} & \eta_{\mathcal{D}}<1 \\ 
\displaystyle \frac{2F_{0}\csc \alpha}{\eta_{\mathcal{R}}}
\left( -r_{1}+r_{2}\eta_{\mathcal{R}}+\left( r_{1}\eta_{\mathcal{R}}
-r_{2}\right) \cos \alpha  +r_{1}\alpha \eta_{\mathcal{R}}\sin \alpha
\right) & \displaystyle \text{for} & \eta_{\mathcal{D}}\geq 1
\end{array}
\right.  \label{Eqn_A_0_41}
\end{equation}
\end{widetext}

\noindent Minimizing this with respect to $\alpha $ we find 
$\alpha^{*}=\mbox{arcsec}(\eta_{\mathcal{R}})$
which gives us the equations 

\begin{equation}
\begin{array}{rl}
W^{*} & \displaystyle = 2F_{0}\left( r_{2}
\sqrt{1-\frac{1}{\eta_{\mathcal{R}}^{2}}}+r_{1}\mbox{arcsec}\left( 
\eta_{\mathcal{R}}\right) \right)  \label{Eqn_A_0_44} \\
\eta_{\mathcal{D}}^{*} & \displaystyle =
\frac{\eta_{\mathcal{R}}\left( r_{1}+r_{2}\eta _{\mathcal{R}}+r_{1}\sqrt{\eta_{\mathcal{R}
}^{2}-1}\,\mbox{arcsec}\left( \eta_{\mathcal{R}}\right) \right) }{\left(
r_{1}\eta_{\mathcal{R}}+r_{2}\right) } \\
N^{*} & \displaystyle =\frac{2}{a_{0}}\left( \frac{r_{1}
\eta_{\mathcal{R}}+r_{2}}{\eta_{\mathcal{R}}
\sqrt{\left(\eta_{\mathcal{R}}\right)^{2}-1}}\right) +1
\end{array}
\end{equation}

\noindent With the known values of the constants, equation \ref{Eqn_A_0_44} becomes 

\begin{equation}
W^{*}\left( \eta _{\mathcal{R}}\right) =24k_{B}T \sqrt{1-\frac{1}{\eta_{\mathcal{R}}^{2}}}
+20k_{B}T  \mbox{arcsec}(\eta_{\mathcal{R}}),
\end{equation}

\noindent which is plotted in figure \ref{Figure_06}. At the physiological value of 
$\eta_{\mathcal{R}}=3/2$ for the RecA system, 
$\alpha^{*}= 0.84, \eta_{\mathcal{D}}^{*} = 2.1, 
N^{*} = 25$, and $W^{*} = 174 k_{B}T$. The probability of the system being in the minimal form of structure $1$ is
on the order of $e^{-174}$. This vanishing probability persists for
$\eta_{\mathcal{D}}=\eta_{\mathcal{D}}^{*}\simeq 1.8$.

For somewhat larger $N$ we consider structure $2$ as shown in figure \ref{Figure_05}$B$. 
The angle $\alpha $ no longer enters the calculation directly,
and we deal with the angle $\phi $. The parameters are subject to the
restrictions $1\leq \eta_{\mathcal{R}}\leq 1.8$, $\eta_{\mathcal{D}}
\leq \eta _{\mathcal{R}}$, and $\xi_{\mathcal{D}}\leq \xi_{\mathcal{R}}$.

The exchange regions can simultaneously be in homologous alignment if $N$
satisfies

\begin{equation}
a=a_{0}\eta_{\mathcal{D}}\left( N-1\right)\, \text{ and }\, b=a_{0}\eta_{\mathcal{R}}\left( N-1\right),
\label{Eqn_A_2_05}
\end{equation}

\noindent from which we find $\eta_{\mathcal{D}}/\eta_{\mathcal{R}}=a/b$.

The radius of curvature $R$ for structure $\mathcal{R}$ is related to the opening
angle $\phi $ by 

\begin{equation}
b=R\phi \text{,}  \label{Eqn_A_2_10}
\end{equation}

\noindent while trigonometry gives 

\begin{equation}
a=2R\sin \left( \frac{\phi }{2}\right)  \label{Eqn_A_2_16}
\end{equation}

\noindent Upon using the above, 

\begin{equation}
\eta_{\mathcal{D}}= \frac{2\eta_{\mathcal{R}} \sin \left( \frac{\phi }{2}\right) }{\phi}
\label{Eqn_A_2_20}
\end{equation}

\noindent Using equations \ref{Eqn_A_2_05} and \ref{Eqn_A_2_10} we also find

\begin{equation}
R=\frac{a_{0}\eta_{\mathcal{R}}\left( N-1\right) }{\phi }  \label{Eqn_A_2_25}
\end{equation}

We can now calculate the work required to form structure $2$. We will vary $%
N $ and $\phi $ to minimize this. We can then vary $\eta_{\mathcal{R}}$ and $%
\xi_{\mathcal{R}}$ (subject to $\eta _{\mathcal{R}}\geq \eta_{\mathcal{D}}$ and $%
\xi_{\mathcal{R}}\geq \xi _{\mathcal{D}}$) to examine their effects on the
system. The work to form structure $2$ comes from three terms: stretching $%
\mathcal{D}$, bending $\mathcal{R}$, and separating $\mathcal{D}$ from $%
\mathcal{R}$ against the non-specific attractive force by which they were
initially aligned.

The force required to stretch $\mathcal{D}$ to $\eta_{\mathcal{D}}$ times its
B-form contour length is approximately 
\begin{equation}
F_{\mathcal{D}}=\left\{ 
\begin{array}{lll}
0 & \text{for} & \eta_{\mathcal{D}}<1.0 \\ 
F_{0} & \text{for} & 1.0\leq \eta_{\mathcal{D}}\leq 1.8 \\ 
\infty & \text{for} & 1.8\leq \eta_{\mathcal{D}}
\end{array}
\right.
\end{equation}
Since $\eta _{\mathcal{D}}\geq 1.8$ is unphysical we impose $F_{\mathcal{D}%
}\left( \eta _{\mathcal{D}}\geq 1.8\right) =\infty $ to ensure that this does
not occur. The work to stretch $\mathcal{D}$ is thus 
\begin{equation}
W_{\mathcal{D}}=\left\{ 
\begin{array}{lll}
0 & \text{for} & \eta_{\mathcal{D}}<1 \\ 
F_{0}\left( \eta_{\mathcal{D}}-1\right) a_{0}\left( N-1\right) & \text{for} & 
1\leq \eta_{\mathcal{D}}\leq 1.8 \\ 
\infty & \text{for} & 1.8\leq \eta_{\mathcal{D}}
\end{array}
\right.
\end{equation}

\noindent Using equation \ref{Eqn_A_2_20} and expressing the regime boundaries in
terms of $\phi $ this becomes (in units of $k_{B}T$)

\begin{widetext}
\begin{equation}
W_{\mathcal{D}}=\left\{ 
\begin{array}{lll}
0 & \text{for} & \left( \frac{\sin \left( \frac{\phi }{2}\right) }{\frac{%
\phi }{2}}\right) \eta_{\mathcal{R}}<1 \\ 
F_{0}\left( \left( \frac{\sin \left( \frac{\phi }{2}%
\right) }{\frac{\phi }{2}}\right) \eta_{\mathcal{R}}-1\right) a_{0}\left(
N-1\right) & \text{for} & 1\leq \eta_{\mathcal{R}}\left( \frac{\sin \left( 
\frac{\phi }{2}\right) }{\frac{\phi }{2}}\right) \leq 1.8 \\ 
\infty & \text{for} & 1.8<\eta_{\mathcal{R}}\left( \frac{\sin \left( \frac{%
\phi }{2}\right) }{\frac{\phi }{2}}\right)
\end{array}
\right.
\end{equation}
\end{widetext}

\noindent The work to bend $\mathcal{R}$ into the circular arc in structure $2$ is 

\begin{equation}
W_{\mathcal{R}}=\frac{1}{2}\kappa _{\mathcal{R}}\int \left( \frac{1}{R\left(
s\right) }\right) ^{2}ds=\frac{1}{2}\kappa_{\mathcal{R}}\left( \frac{1}{R}%
\right) ^{2}b
\end{equation}
Using equations \ref{Eqn_A_2_10} and \ref{Eqn_A_2_25} and the fact that $%
\kappa _{\mathcal{R}}\simeq \xi _{\mathcal{R}}$ (in units of $k_{B}T$),
gives 
\begin{equation}
W_{\mathcal{R}}=\frac{\xi_{\mathcal{R}}}{2}\left( \frac{\phi ^{2}}{a_{0}\eta_{%
\mathcal{R}}\left( N-1\right) }\right)
\end{equation}

There is a non-specific attractive interaction between $\mathcal{D}$ and 
$\mathcal{R}$. The work to pull $\mathcal{D}$ and $\mathcal{R}$ apart is
approximately proportional to the length of $\mathcal{R}$ between the
exchange regions. 
For intermediate $N$ we ignore
an energetic contribution $\propto \varepsilon k_{B}T a_{0}\eta_{\mathcal{R}}N$,
which underestimates the work and produces a lower bound,

\begin{equation}
W(\phi)=\left\{ 
\begin{array}{lll}
W_{\mathcal{R}} & \text{for} & \left( \frac{\sin \left( \frac{\phi }{2}\right) }{\frac{%
\phi }{2}}\right) \eta _{\mathcal{R}}<1 \\ 
W_{\mathcal{D}}+W_{\mathcal{R}}  & \text{for} & 1\leq \eta _{\mathcal{R}}\left( \frac{\sin \left( \frac{%
\phi }{2}\right) }{\frac{\phi }{2}}\right) \leq 1.8 \\ 
\infty & \text{for} & 1.8<\eta _{\mathcal{R}}\left( \frac{\sin \left( \frac{%
\phi }{2}\right) }{\frac{\phi }{2}}\right)
\end{array}
\right.
\end{equation}

Upon minimizing the work $W(\phi)$ with respect to the angle $\phi$, we find $\phi^{*}$ and 

\begin{widetext}
\begin{equation}
\begin{array}{lr}
W(\phi^{*}) \equiv W^{*} & \displaystyle = a_{0}F_{0}\left( N-1\right) \eta _{\mathcal{R}}
\left\{ \left( \frac{\eta _{\mathcal{R}}-1}{\eta _{\mathcal{R}}}\right) \right. 
+ \left( \frac{1}{120}\max \left[ 0,\min \left[ 10-2\sqrt{\frac{30}
{\eta_{\mathcal{R}}}-5},10-\frac{120k_{B}T\xi _{\mathcal{R}}}{a_{0}^{2}F_{0}\left(
N-1\right) ^{2}\eta _{\mathcal{R}}^{2}}\right] \right] \right.  \\[13pt]
\:&\displaystyle \times\left( \frac{240k_{B}T\xi _{\mathcal{R}}}{a_{0}^{2}F_{0}\left(
N-1\right) ^{2}\eta _{\mathcal{R}}^{2}}-20\right.\left. \left. \left. \max \left[ 0,\min \left[ 10-2\sqrt{\frac{30}{\eta _{\mathcal{R}}}-5},
10-\frac{120k_{B}T\xi _{\mathcal{R}}}{a_{0}^{2}F_{0}\left(
N-1\right) ^{2}\eta _{\mathcal{R}}^{2}}\right] \right] \right) \right) \right\} \:\hspace{1cm}\:
\end{array}
\end{equation}
\end{widetext}

\noindent This function is plotted in figures \ref{Figure_07}$A$ and $B$ using standard values for $a_{0}, F_{0}$, 
and $k_{B}T$. From figures \ref{Figure_07}$A$ and $B$ it is clear that by $N\approx 60$ the
minimal form of structure $2$ is dominated by bending $\mathcal{R}$. For
large $N$, $\left( \gtrsim 60\right) $, the interaction energy can no longer
be ignored, but now we always have $\left( \sin \left( \frac{\phi }{2}
\right) /\frac{\phi }{2}\right) \eta_{\mathcal{R}}<1$. The total work thus
simplifies to 

\begin{equation}
W=\frac{4k_{B}T\left( 5-\sqrt{\frac{30}{\eta _{\mathcal{R}}}-5}\right) \xi _{\mathcal{R}}
}{a_{0}\left( N-1\right) \eta _{\mathcal{R}}}+\varepsilon k_{B}T a_{0}\eta _{\mathcal{R}
}\left( N-1\right)   
\label{Eqn_A_3_09}
\end{equation}

\noindent The separation between exchange regions which minimizes this and the
corresponding minimum work are 

\begin{equation}
N^{*}=\frac{2}{a_{0}\eta _{\mathcal{R}}}\sqrt{\left( \frac{\xi _{\mathcal{R}}}{
\varepsilon }\right) \left( 5-\sqrt{\frac{30}{\eta _{\mathcal{R}}}-5}\right) }
\label{Eqn_A_3_10}
\end{equation}
and 
\begin{equation}
W^{*}=4k_{B}T\sqrt{\left( \xi _{\mathcal{R}}\varepsilon \right) \left( 5-
\sqrt{\frac{30}{\eta _{\mathcal{R}}}-5}\right) }.
\end{equation}

\noindent These are plotted in figures \ref{Figure_08} and \ref{Figure_09}
respectively. We also note that $N^{*}$ is monotonic with respect to the 
$\xi _{\mathcal{R}}$ and $\varepsilon $ but has a maximum with respect to 
$\eta_{\mathcal{R}}$ at $\eta _{\mathcal{R}}^{*}=\frac{5}{4}\left( 3+\sqrt{3}\right)
\simeq 1.58$.

\end{appendix}

\end{document}